\documentclass[prl,showpacs,superscriptaddress,
twocolumn,
nofootinbib,
floatfix]{revtex4-2}

\usepackage[english]{babel}
\usepackage[utf8]{inputenc}
\usepackage{multirow}
\usepackage{amsmath,amssymb}
\usepackage{bbold}
\usepackage{slashed}
\usepackage{subfigure}
\usepackage[colorlinks=true,
breaklinks=true,
urlcolor=magenta,
citecolor=blue]{hyperref}
\usepackage[usenames,dvipsnames]{color}
\usepackage{appendix}
\usepackage{braket,bm}
\usepackage{multirow}
\usepackage{enumitem}
\usepackage{color}
\usepackage{cancel}
\usepackage{dsfont}
\usepackage{orcidlink}
\usepackage{mathrsfs}
\usepackage{latexsym}
\usepackage{graphicx}
\usepackage{indentfirst}
\usepackage{bbm}
\usepackage{tabularx}

\graphicspath{{figs/}}

\newcommand{\be}{\begin{equation}} \newcommand{\ee}{\end{equation}}
\newcommand{\ba}{\begin{array}{c}} \newcommand{\ea}{\end{array}}
\newcommand{\bea}{\begin{eqnarray}} \newcommand{\eea}{\end{eqnarray}}

\allowdisplaybreaks

\newcommand{\itp}
{\affiliation{Institute of Theoretical Physics, Chinese Academy of Sciences, Beijing 100190, China}}

\newcommand{\ucas}{\affiliation{School of Physical Sciences, University of Chinese Academy of Sciences, Beijing 100049, China}}

\newcommand{\hbnu}{\affiliation{College of Physics and Hebei Key Laboratory of Photophysics Research and Application,\\ Hebei Normal University, Shijiazhuang 050024, China}}

\newcommand{\hnu}{\affiliation{School of Physics and Electronics, Hunan University, Changsha 410082, China}}

\newcommand{\huHESPA}{\affiliation{Hunan Provincial Key Laboratory of High-Energy Scale Physics and Applications,\\ Hunan University, Changsha 410082, China}}

\begin{document}
\title{Two-Loop Extraction of the Pion-Nucleon Sigma Term}

\author{Ze-Rui~Liang\orcidlink{0009-0007-1311-1796}}
\hbnu\hnu
\author{Han-Xue~Chen\orcidlink{0009-0006-8886-0241}}
\hnu\huHESPA
\author{Feng-Kun~Guo\orcidlink{0000-0002-2919-2064}}
\email{fkguo@itp.ac.cn}
\itp\ucas
\author{Zhi-Hui~Guo\orcidlink{0000-0003-0409-1504}}
\hbnu
\author{De-Liang~Yao\orcidlink{0000-0002-5524-2117}}
\email{yaodeliang@hnu.edu.cn}
\hnu\huHESPA

\begin{abstract}

The pion-nucleon sigma term, characterizing the mass component of Higgs origin related to $u$ and $d$ quarks inside the nucleon, is investigated within relativistic baryon chiral perturbation theory at leading two-loop order using the extended-on-mass-shell renormalization scheme. 
The two-loop representation of the sigma term is derived from the nucleon mass via the Feynman-Hellmann theorem and verified through a direct calculation of the forward isoscalar-scalar nucleon matrix element. We apply the derived chiral expression to extract the physical pion-nucleon sigma term by extrapolating $N_f=2+1$ lattice quantum chromodynamics (QCD) data at unphysical quark masses. 
We find that, at the two-loop level, the long-standing tension between lattice QCD and dispersive determinations can be naturally resolved, owing to the incorporation of intermediate $\pi\pi$ rescattering effects that begin to contribute at two-loop order. 
Our final result for the nucleon sigma term based on recent lattice QCD calculations is $\sigma_{\pi N}=55.9(2.5)$~MeV. It is compatible with the result of the Roy-Steiner equation analysis and thus provides a satisfactory resolution to the previous debate between lattice QCD and phenomenological determinations.

\end{abstract}
\date{\today}

\maketitle

{\it Introduction---}The pion-nucleon sigma term is defined by the nucleon matrix element of the isoscalar scalar quark current at zero momentum transfer~\cite{Reya:1974gk}
\begin{align}
\sigma_{\pi N}=\frac{1}{2m_N}\langle N|\hat{m}(\bar{u}u+\bar{d}d)|N\rangle \, , \label{eq.sigma.def0}
\end{align}
with $\hat{m}=(m_u+m_d)/2$, characterizing the portion of nucleon mass due to nonvanishing up and down quark masses, which are of Higgs origin, and encoding the underlying information on the scalar couplings of the nucleon to these quarks. 
Knowledge of $\sigma_{\pi N}$ is of vital importance in particle and nuclear physics, as it serves as essential input for calculations across a variety of significant topics including proton mass decomposition~\cite{Ji:1994av}, scalar dark matter detection~\cite{Bottino:1999ei,Bottino:2001dj,Ellis:2008hf}, nucleosynthesis~\cite{Berengut:2013nh}, etc. 

Phenomenological extraction of $\sigma_{\pi N}$ relies on the so-called Cheng-Dashen (CD) low-energy theorem~\cite{Cheng:1970mx,Brown:1971pn}, which relates $\sigma_{\pi N}$ directly to the isospin-even ${\pi N}$ scattering amplitude at the subthreshold CD point. A pioneering work using the experimental ${\pi N}$ data available at that time led to the prevalent result of $\sigma_{\pi N}\sim 45$~MeV~\cite{Gasser:1990ce}.
The modern Roy-Steiner (RS) equation analysis of ${\pi N}$ scattering with constraints from precise data of pionic hydrogen and deuterium shifts the sigma term upward to $\sigma_{\pi N}= 59.1(3.5)$~MeV~\cite{Hoferichter:2015dsa}. 
This value is robust in the sense that axiomatic $S$-matrix principles such as unitarity, analyticity, and crossing symmetry have been implemented to extrapolate the ${\pi N}$ amplitude to the unphysical CD point in a reliable manner. 
It is also pointed out in Ref.~\cite{Hoferichter:2023ptl} that isospin-breaking effects cause only a small change of $\Delta \sigma_{\pi N}=3.5~\rm{MeV}$.

However, a long-standing tension persists between the RS determination and lattice quantum chromodynamics (QCD) extractions. As summarized in the review by the Flavour Lattice Averaging Group (FLAG)~\cite{FlavourLatticeAveragingGroupFLAG:2024oxs}, lattice calculations for $\sigma_{\pi N}$ using ensembles with $N_f=2+1$ dynamical quarks have mostly yielded values smaller than $50$~MeV~\cite{Durr:2015dna,Yang:2015uis,RQCD:2022xux,Agadjanov:2023efe}, which are consistent with the pioneering estimate~\cite{Gasser:1990ce} but disagree with the modern RS result~\cite{Hoferichter:2015dsa}. 
The origin of this discrepancy remains unclear and poses a significant challenge for precision strong interaction physics; see Refs.~\cite{Alarcon:2021dlz, Hoferichter:2025ubp} for recent reviews. 

In lattice QCD, the sigma term can be calculated using two strategies: one is to directly compute the three-point scalar matrix element in Eq.~\eqref{eq.sigma.def0}, and the other is to derive it from the quark-mass derivative of the nucleon mass by applying the Feynman-Hellmann (FH) theorem~\cite{Hellmann:1937book,Feynman:1939zza}. 
Based on the former approach, it is shown in Ref.~\cite{Gupta:2021ahb} that the discrepancy can be remedied by properly considering the excited-state contamination (ESC)~\cite{Bar:2016uoj} using baryon chiral perturbation theory (BChPT)~\cite{Weinberg:1978kz,Gasser:1983yg,Gasser:1984gg,Gasser:1987rb,Bernard:1995dp,Scherer:2012xha}, where a ten-MeV enhancement can be obtained, leading to $\sigma_{\pi N}=59.6(7.4)$~MeV. 
However, in a recent lattice calculation where the excited-state contribution is taken into account~\cite{Agadjanov:2023efe}, an upward trend was observed but not as pronounced. Thus, it remains unclear whether the subtraction of ESC plays the sole role in alleviating the tension.

In fact, apart from the ESC, the chiral extrapolation of lattice QCD data for $\sigma_{\pi N}$ may also be problematic. 
In particular, isoscalar scalar currents couple to the lightest scalar meson, the $\sigma$ (also known as the $f_0(500)$), which becomes a bound state below the two-pion threshold at unphysical pion masses around $300$~MeV~\cite{Briceno:2016mjc,Briceno:2017qmb,Cao:2023ntr}. 
Even though the method relying on the FH theorem does not require external scalar currents, in the chiral expansion of the nucleon mass, the two pions in the pionic loops can still couple to the $\sigma$ meson. Furthermore, the $\pi\pi$ scattering is part of the RS equation analysis~\cite{Hoferichter:2015tha}. These observations suggest that the rescattering effects of intermediate isoscalar-scalar $\pi\pi$ pairs could be crucial for extracting the correct $\sigma_{\pi N}$ at unphysical pion masses. To incorporate such effects, higher-order calculations of $\sigma_{\pi N}$ beyond one loop in BChPT are required. However, to date, expressions used for extrapolation are either limited to the one-loop level~\cite{Alvarez-Ruso:2013fza,Ren:2017fbv,Agadjanov:2023efe} or are merely a polynomial in $M_\pi$~\cite{Yang:2015uis}. 

In this Letter, we solve this problem by deriving the leading two-loop chiral expression of $\sigma_{\pi N}$ in relativistic BChPT~\cite{Gasser:1987rb,Bernard:1995dp,Scherer:2012xha} using the extended-on-mass-shell (EOMS) scheme~\cite{Fuchs:2003qc}. 
The obtained two-loop EOMS expression is renormalization-scale independent, exhibits correct power counting, and respects the analyticity of relativistic field theory, thereby enabling a reliable and robust extraction of $\sigma_{\pi N}$ from lattice QCD data via chiral extrapolation.
As will be shown below, the $\sigma_{\pi N}$ obtained from two-loop chiral extrapolation of the lattice data in Ref.~\cite{Agadjanov:2023efe} is consistent with the RS result, thus resolving the remaining tension between lattice QCD~\cite{Agadjanov:2023efe} and phenomenological determinations~\cite{Hoferichter:2015dsa}.

{\it Pion-nucleon sigma term at leading two-loop order---}In BChPT, the sigma term $\sigma_{\pi N}$ can be derived either via the nucleon scalar form factor at $q^2=0$, with $q$ denoting the four-momentum transferred between the initial and final nucleons, or by applying the FH theorem~\cite{Hellmann:1937book,Feynman:1939zza} to the chiral expression of the nucleon mass. The two approaches are equivalent and will be used simultaneously for a consistency check. Specifically, $\sigma_{\pi N}$ is associated with the derivative of the nucleon mass with respect to the leading-order pion mass $M$, i.e.,
\begin{align}
\sigma_{\pi N}=\hat{m}\frac{\partial m_N}{\partial \hat{m}}= M^2\frac{\partial m_N }{\partial M^2} \, .
\end{align}
The second equality holds by virtue of the Gell-Mann-Oakes-Renner relation~\cite{Gell-Mann:1968hlm}, $M^2=2B_0 \hat{m}$ (where $\hat{m}=(m_u+m_d)/2$), and $B_0$ is a constant related to the quark condensate through $B_0=-\langle\bar{q}q\rangle/(3F^2)$ with $F$ being the pion decay constant in the chiral limit.

\begin{figure}[tbh]
    \centering
    \includegraphics[width=\linewidth, angle=-0]{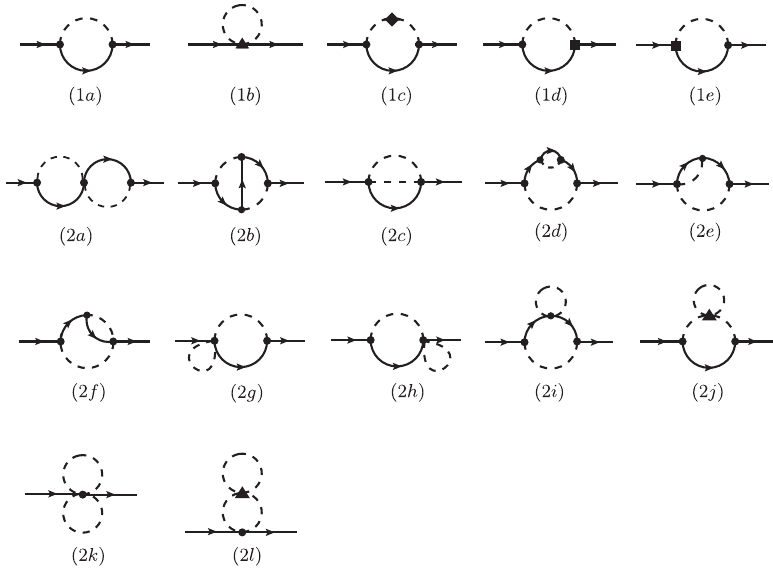}
    \caption{One-loop and two-loop Feynman diagrams contributing to the nucleon mass and $\pi N$ sigma term $\sigma_{\pi N}$. The solid and dashed lines denote nucleons and pions, respectively.
    The dots, triangles, squares and diamond represent the vertices of $\mathcal{O}(p^1)$, $\mathcal{O}(p^2)$, $\mathcal{O}(p^3)$ and $\mathcal{O}(p^4)$, in order.}
    \label{fig.loops_onecol}
\end{figure}
The chiral expression for the nucleon mass has been obtained in BChPT up to $\mathcal{O}(p^5)$~\cite{Liang:2025cjd}, i.e., the leading two-loop order, where the feasibility of the EOMS scheme at the two-loop level is verified using the method of dimensional counting analysis~\cite{Gegelia:1999qt} or equivalently the method of regions~\cite{Smirnov:2002pj}.\footnote{Although there exist other two-loop calculations of the nucleon mass in Refs.~\cite{McGovern:1998tm,Schindler:2006ha,Schindler:2007dr,Chen:2024twu}, these results do not keep the full analytic structures.} The resulting nucleon mass in BChPT can be written as
\begin{align}\label{eq.chpt.mn.tlo.full}
    m_N = \tilde{m}-4\tilde{c}_1 M^2-2\tilde{e}_m M^4 + 
    \Delta m^{(1)}+
    \Delta m^{(2)}\, ,
\end{align}
where $\Delta m^{(1)}$ and $\Delta m^{(2)}$ denote contributions from one- and two-loop Feynman diagrams, shown as (1a, $\ldots$, 1e) and (2a, $\ldots$, 2l) in Fig.~\ref{fig.loops_onecol}, respectively. 
The parameters with tildes, $\tilde{m}$, $\tilde{c}_1$ and $\tilde{e}_m$, represent renormalized LECs 
using the EOMS scheme~\cite{Fuchs:2003qc}, and their dependence on the renormalization scale $\mu$ is given by
\begin{align}
    X^r = \widetilde{X} + \delta^{(21)}_X +  \delta^{(22)}_X\, ,\quad X\in\{m, c_1, e_m\}\, ,
\end{align}
and the renormalized LECs $X^r$ are defined by
\begin{align}\label{eq.LECs.UV.tlo}
X=X^r+\frac{\beta_X^{(22)}}{\epsilon^2\Lambda^2}
-\frac{\beta_X^{(21)}}{\epsilon\Lambda^2}-\frac{\beta_X^{(11)}}{\epsilon\Lambda}\, ,
\end{align}
with $\Lambda=1/(16\pi^2)$, $\epsilon=(4-d)/2$ where $d$ is the spacetime dimension, $\delta^{(ij)}_X$ denoting the finite shifts that remedy the power counting breaking issue, and $\beta_X^{(ij)}$ representing the ultraviolet (UV) beta functions whose explicit expressions are given in Ref.~\cite{Liang:2025cjd}.

With the two-loop nucleon mass in Eq.~\eqref{eq.chpt.mn.tlo.full}, the chiral result for $\sigma_{\pi N}$ is derived as
\begin{align}\label{eq.simga.term.full}
    \sigma_{\pi N} = -4\tilde{c}_1 M^2-4\tilde{e}_m M^4 + \Delta \sigma_{\pi N}^{(1)}+
   \Delta \sigma_{\pi N}^{(2)},
\end{align}
where the contributions from the one- and two-loop Feynman diagrams are obtained from $\Delta m^{(n)}$ via $\Delta\sigma_{\pi N}^{(n)}\equiv M^2({\partial \Delta m^{(n)}}/{\partial M^2})$ with $n\in\{1,2\}$. Explicit chiral expressions for $\Delta \sigma_{\pi N}^{(1)}$ and $\Delta \sigma_{\pi N}^{(2)}$ are given in the Supplementary Material.
We have verified these expressions by directly calculating the scalar nucleon matrix element at zero-momentum transfer, using a procedure similar to the two-loop calculation of $g_A$ in Ref.~\cite{Bernard:2025gto}.

{\it Pion mass dependence---}Lattice QCD calculations are typically performed at unphysical quark masses, making chiral extrapolation to the physical point necessary. The chiral expression for $\sigma_{\pi N}$ obtained at $\mathcal{O}(p^5)$ within the EOMS scheme is well-suited for conducting chiral extrapolation, as the uncertainty due to truncation of higher-order corrections should be much smaller than that of the one-loop expression used in, e.g., Ref.~\cite{Agadjanov:2023efe}. Additionally, intermediate $\pi\pi$ rescattering effects begin to contribute at leading two-loop order, which are absent at one loop.

Recently, a lattice QCD analysis of the nucleon sigma term has been carried out by calculating the three-point scalar matrix element using $N_f=2+1$ dynamical fermions. 
Although a comprehensive error analysis, including chiral extrapolation, ESC, finite volume (FV) and lattice spacing effects, has been accomplished, a small value of $\sigma_{\pi N}=43.7(3.6)$~MeV was still obtained~\cite{Agadjanov:2023efe}. 
In obtaining that value, an SU(3) $\mathcal{O}(p^3)$ (one-loop order) chiral expression for $\sigma_{\pi N}$~\cite{Lehnhart:2004vi,Severt:2019sbz} was employed. 
Here, we will show that when the leading two-loop SU(2) $\mathcal{O}(p^5)$ result,\footnote{It has been demonstrated in Refs.~\cite{Alvarez-Ruso:2013fza,Ren:2016aeo} that SU(2) BChPT can be applied to study $N_f = 2 + 1$ lattice QCD simulations as long as the strange-quark mass is close to its physical value.} as given in Eq.~\eqref{eq.simga.term.full}, is adopted, a larger value that is consistent with the RS equation result~\cite{Hoferichter:2015dsa} can be achieved. 
In contrast, if we perform the fit with the $\mathcal{O}(p^3)$ expression, as discussed in detail in the Supplementary Material, one would indeed get a smaller value for $\sigma_{\pi N}$. 

To perform chiral extrapolation of lattice QCD data~\cite{Agadjanov:2023efe} with our chiral expression, the following merit function is employed in fitting to the lattice data:
\begin{align}
\chi^2 = \chi^2_{m_N}+\omega\cdot(\chi^2_{\rm win} + \chi^2_{\rm sum} + \chi^2_{\rm exp})\, ,
\end{align}
with $\chi^2_{m_N}=(m_N^{\rm ChPT}-m_N^{\rm phy.})^2/(\delta m_N^{\rm phy.})^2$ imposing the constraint of the physical nucleon mass with $m_N^{\rm phy.}=938.92$~MeV and $\delta m_N^{\rm phy.}=0.65$~MeV~\cite{ParticleDataGroup:2024cfk}, where the central value corresponds to the average mass of the proton and neutron, and the uncertainty is taken as half of their mass difference. Three approaches are employed in Ref.~\cite{Agadjanov:2023efe} to address the excited-state contribution of the nucleon correlation function. 
We include all these three datasets with the same weight $\omega=1/3$ since they are not independent of each other. The datasets are labeled with the subscripts ``win'', ``sum'', and ``exp'' for the window average of the summed correlator, the explicit two-state fit to the summed correlator, and the explicit two-state fit to the effective form factor, respectively; see Ref.~\cite{Agadjanov:2023efe} for further details. 

Lattice calculations are performed at discretized space points with lattice spacing $a$ in a FV of size $L^3$. 
The FV and lattice spacing corrections can be included as:
\begin{align}
    \sigma_{\pi N}+ \Delta_L \sigma_{\pi N} + b_\pi \frac{a}{\sqrt{t_0}}M_\pi^2,
\label{eq:sigma_FVC}
\end{align}
where the FV correction $\Delta_L \sigma_{\pi N} = \Delta_L \sigma_{\pi N}^{(N)} (M_\pi;L) + \Delta_L \sigma_{\pi N}^{(\Delta)} (M_\pi;L)$ is estimated using one-loop BChPT with explicit $\Delta$ resonances in a FV~\cite{Beane:2004tw,Geng:2011wq,Ren:2012aj,Alvarez-Ruso:2013fza,Liang:2022tcj} (see the Supplementary Material), and the last term accounts for the fact that the current used in Ref.~\cite{Agadjanov:2023efe} is not $\mathcal{O}(a)$ improved. Note that strict ChPT calculations of FVC at two-loop level have only been realized for the meson sector~\cite{Colangelo:2006mp,Bijnens:2014dea} so far, and the results indicate that the two-loop contribution is suppressed compared to the leading-loop one. The asymptotic behavior of $\Delta_L \sigma_{\pi N}$ coincides with that of Ref.~\cite{Beane:2004tw}, which decreases exponentially with $M_\pi L$ as $L\to\infty$. In Eq.~\eqref{eq:sigma_FVC}, the lattice spacing parameter $b_\pi$ is unknown, and the gradient flow scale $\sqrt{t_0}$~\cite{Bruno:2016plf} is taken as $0.14464(87)$~fm at the physical point~\cite{FlavourLatticeAveragingGroupFLAG:2021npn}.

\begin{table}[tb]
\setlength{\arraycolsep}{0.1pt}
\renewcommand{\arraystretch}{1.3}
\caption{Fit results of the two-loop renormalized LECs.
}\label{tab:fit.LECs}
\begin{tabular}{lcccc}
\hline\hline
\multicolumn{1}{c}{\multirow{2}*{LECs}} & \multicolumn{1}{c}{\multirow{2}*{Values}}   &\multicolumn{3}{c}{ Correlation matrix} \\
\cline{3-5}
 &  & $\tilde{m}$ & $\tilde{c}_1$   & $\tilde{e}_m$\\
\hline
$\tilde{m}$~ [MeV]  & $863.9\pm 2.1$  &$1.000$  &$0.944$ &$-0.662$ \\
$\tilde{c}_1$~ [GeV$^{-1}$] & $-1.06\pm0.02$  &   & $1.000$  &$-0.728$\\
$\tilde{e}_m$ [GeV$^{-3}$] & $-5.59\pm 0.21$ &   &  & $1.000$ \\
\hline
${\chi^2}/{\rm d.o.f.}$  & {${18.24}/{(15+1-4)}\simeq 1.52$}& &\\
\hline\hline
\end{tabular}
\end{table}

We treat the two-loop renormalized LECs $\tilde{X}\in\{\tilde{m},\tilde{c}_1,\tilde{e}_m\}$ and $b_\pi$ as free parameters, while the remaining ones are fixed following Ref.~\cite{Liang:2025cjd,Yao:2016vbz,Yao:2017fym}. 
The fit includes lattice data points for $M_\pi$ from 154 to 352~MeV and excludes the one at $M_\pi=128$~MeV, since our primary aim is to predict the value of $\sigma_{\pi N}$ at the physical pion mass. The results are compiled in Table~\ref{tab:fit.LECs}. 
The obtained value of $\tilde{c}_1$ agrees marginally with that determined by matching $\pi N$ RS equations to BChPT at $\mathcal{O}(p^4)$, $c_1^{\rm RS} = -1.11(3)$~GeV$^{-1}$~\cite{Hoferichter:2015tha}. For the lattice spacing parameter, we obtain $b_\pi=0.20(20)$~GeV$^{-1}$.

\begin{figure}[tb]
\centering
\includegraphics[width=0.98\linewidth]{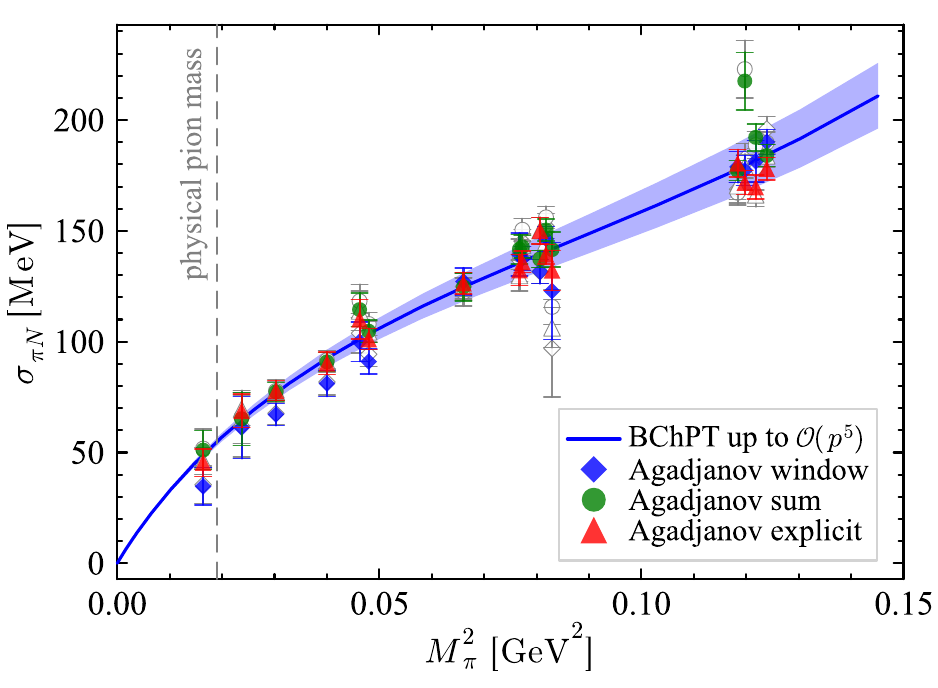}
\caption{Pion mass dependence of the pion-nucleon sigma term $\sigma_{\pi N}$. The lattice QCD data from Ref.~\cite{Agadjanov:2023efe} are shifted to the infinite-volume continuum results by subtracting the FV and lattice spacing corrections. The shifted data are represented by red upper triangles (explicit), green filled circles (sum), and blue diamonds (window), while the unshifted data are displayed using the corresponding gray markers. The dashed vertical line denotes the isospin averaged physical pion mass $M_\pi=138$~MeV, below which the data points are excluded in the fit.}
\label{fig:sigma.Aga}
\end{figure}

\begin{figure}[t]
\centering
\includegraphics[width=\linewidth]{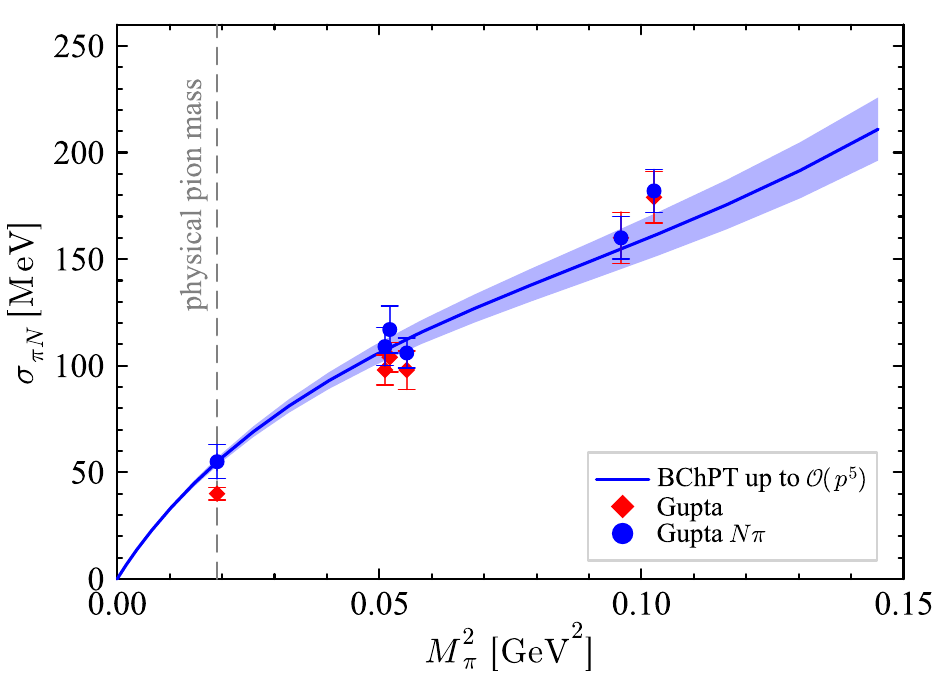}
\caption{Comparison of our prediction for the pion-nucleon sigma term $\sigma_{\pi N}$ with lattice QCD results from Ref.~\cite{Gupta:2021ahb}. Here, blue filled circles and red diamonds denote the lattice results with and without considering the ESC from $N\pi(\pi)$ states, respectively.}
\label{fig:sigma.Gupta}
\end{figure}

In Fig.~\ref{fig:sigma.Aga}, we show the pion mass dependence of $\sigma_{\pi N}$ from the best fit with the leading two-loop BChPT expression, along with the lattice QCD data~\cite{Agadjanov:2023efe}. 
The agreement is excellent. The fit slightly favors the ``exp" dataset over the other two datasets, as indicated by $\{\chi^2_{\rm win},\chi^2_{\rm sum},\chi^2_{\rm exp}\}/N_{\rm data}=\{1.22,1.40,1.04\}$ with $N_{\rm data}=15$. In Ref.~\cite{Gupta:2021ahb}, the contributions of $N\pi$ and $N\pi\pi$ intermediate states to the three-point nucleon correlation function are computed in BChPT to quantify the size of ESC. 
It is therefore interesting to compare our two-loop $\sigma_{\pi N}$ with the lattice QCD data provided therein, as shown in Fig.~\ref{fig:sigma.Gupta}. 
We find that our result demonstrates remarkable consistency with the dataset in which the ESC is properly accounted for.

{\it Extraction of the sigma term---}With the parameters in Table~\ref{tab:fit.LECs}, we predict $\sigma_{\pi N}$ in the physical world,
\begin{align}
\sigma_{\pi N}&=55.9\pm(1.9)_{{\rm stat}}\pm(1.5)_{{\rm sys}_1}\pm( 0.6)_{{\rm sys}_2}\rm{MeV} ,
\end{align}
where the physical pion mass is taken to be the charged one, following the convention in Refs.~\cite{ Meissner:1997ii, Hoferichter:2015dsa}. The uncertainties in $\sigma_{\pi N}$ stem from three sources: (i) statistical (stat) uncertainty propagated from the $1\sigma$ errors of the fitted parameters ($\tilde{m}$, $\tilde{c}_1$, and $\tilde{e}_m$); (ii) systematic uncertainty (sys$_1$) arising from the errors in the one-loop renormalized LECs, $c_2=3.35(3)$ and $c_3=-5.70(6)$~\cite{Siemens:2016jwj}, involved in the $\mathcal{O}(p^4)$ chiral loop contribution; (iii) uncertainty (sys$_2$) due to truncation of the chiral expansion at $\mathcal{O}(p^5)$ and estimated using the method proposed in Ref.~\cite{Epelbaum:2014efa}, which is consistent with the Bayesian method in Refs.~\cite{Schindler:2008fh,Furnstahl:2015rha}.

\begin{figure}[t!]
\centering
\includegraphics[width=0.99\linewidth]{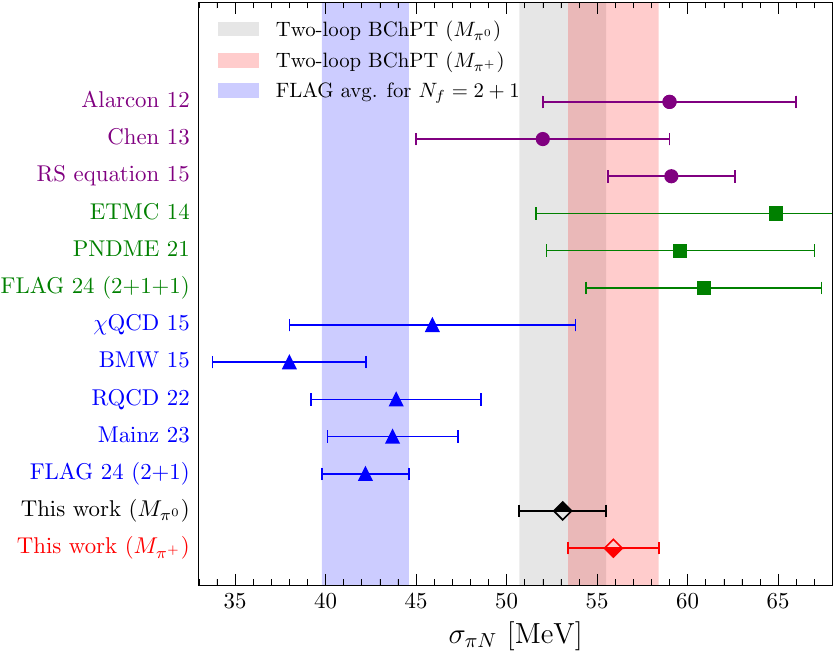}
\caption{Comparison of our result to other determinations. Filled purple circles denote phenomenological determinations: 
Alarcon 12~\cite{Alarcon:2011zs}, 
Chen 13~\cite{Chen:2012nx}, 
RS equation 15~\cite{Hoferichter:2015dsa}; 
green squares stand for the $N_f=2+1+1$ lattice QCD results and the FLAG average: ETMC 14~\cite{Alexandrou:2014sha}, PNDME 21~\cite{Gupta:2021ahb} and FLAG 24 (2+1+1)~\cite{FlavourLatticeAveragingGroupFLAG:2024oxs}; and blue triangles represent 
$N_f=2+1$ lattice QCD results and the FLAG average: 
$\chi$QCD 15~\cite{Yang:2015uis}, 
BMW 15~\cite{Durr:2015dna}, RQCD 22~\cite{RQCD:2022xux}, Mainz 23~\cite{Agadjanov:2023efe}, and
FLAG 24 (2+1)~\cite{FlavourLatticeAveragingGroupFLAG:2024oxs}.
}
\label{fig:comparison}
\end{figure}

Combining the three uncertainties in quadrature, we obtain the final result:
\begin{align}
 \sigma_{\pi N} &=55.9(2.5)~\rm{MeV}\, ,
\end{align}
representing the first extraction of the nucleon sigma term using the complete EOMS chiral expression at the leading two-loop level. This result is consistent with the dispersive determination $\sigma_{\pi N}=59.1(3.5)$~MeV within uncertainties, resolving the tension between lattice QCD and phenomenology. Figure~\ref{fig:comparison} compares our $\sigma_{\pi N}$ result with both lattice QCD and phenomenological determinations,\footnote{The ETMC $N_f=2+1+1$ results~\cite{Alexandrou:2019brg,Alexandrou:2024ozj} disagree with the FLAG $N_f=2+1+1$ average. The reason is still unclear and hence we do not show them in Fig.~\ref{fig:comparison}.} where our result with the physical pion mass taking the value of $M_{\pi^0}$, $\bar{\sigma}_{\pi N}=53.1(2.4)$~MeV, is exhibited as well.

It should be emphasized that a leading one-loop [$\mathcal{O}(p^3)$] analysis of the same dataset from Ref.~\cite{Agadjanov:2023efe} yields a smaller value of $\sigma_{\pi N}=48.1(1.9)$~MeV. The incorporation of two-loop effects shifts $\sigma_{\pi N}$ upward by about 8~MeV. 
This enhancement indicates the importance of isoscalar $\pi\pi$ rescattering which enters at the two-loop level, exemplified by diagram ($2j$) in Fig.~\ref{fig.loops_onecol}, which is absent at the one-loop level. Among all the $\mathcal{O}(p^5)$ two-loop diagrams, this diagram generates the largest positive contribution to the sigma term, amounting to $4.0$~MeV with the renormalization scale taken at the nucleon mass in the chiral limit. We find that using a complete one-loop [$\mathcal{O}(p^4)$] BChPT expression can also yield a large $\sigma_{\pi N}$ of $56.7(1.9)$~MeV, as shown in the Supplementary Material, albeit with a different value of the $\mathcal{O}(p^4)$ LEC $\tilde e_m$ from that in Table~\ref{tab:fit.LECs}. 
This finding can be explained by the fact that the $e_m$ term mimics part of the two-loop contribution through parameter adjustment.

{\it Conclusion---}We have calculated the pion-nucleon sigma term within the framework of covariant BChPT up to $\mathcal{O}(p^5)$, i.e., the leading two-loop order, using the EOMS renormalization scheme. 
The resulting chiral expression maintains correct power counting, preserves the original analyticity, is renormalization scale-independent, and partially incorporates the $\pi\pi$ rescattering effect. 
The application of the leading two-loop BChPT expression to the chiral extrapolation of $N_f=2+1$ lattice QCD data~\cite{Agadjanov:2023efe} leads to a $\sigma_{\pi N}$ value consistent with the dispersive determination~\cite{Hoferichter:2015dsa}, providing a natural resolution to the discrepancy between lattice QCD and phenomenological analyses that has persisted over the past decade. 
With leading-two-loop-order precision, our result for the nucleon sigma term is $\sigma_{\pi N} =55.9(2.5)~\rm{MeV}$, favoring the larger-value solution for $\sigma_{\pi N}$.
Our finding demonstrates the general importance of $\pi\pi$ rescattering effects in observables coupled to isoscalar scalar currents, even though such contributions only emerge at the two-loop level.

\medskip

\begin{acknowledgements}
{\bf Acknowledgements:} We would like to thank Ulf-G. Mei{\ss}ner for helpful discussions. DLY and ZRL appreciate the hospitality of Institute of Theoretical Physics (ITP) at Chinese Academy of Sciences (CAS), where part of this work was done. This work is supported by National Nature Science Foundations of China under Grants No.~12275076, No.~11905258, No.~12335002, No.~12475078, No.~12125507, and No.~12447101; by the Science Fund for Distinguished Young Scholars of Hunan Province under Grant No.~2024JJ2007; by the Fundamental Research Funds for the Central Universities under Grant No.~531118010379; by the Science Foundation of Hebei Normal University under Grants No. L2025B09 and No. L2023B09; by Science Research Project of Hebei Education Department under Grant No. QN2025063; by Hebei Natural Science Foundation under Grant No. A2025205018; by the National Key R\&D Program of China under Grant No. 2023YFA1606703; and by CAS under Grant No. YSBR-101.
\end{acknowledgements}

\bibliography{refs}
\setcounter{section}{0}
\renewcommand{\thesection}{S\arabic{section}}

\setcounter{equation}{0}
\renewcommand{\theequation}{S\arabic{equation}}

\setcounter{figure}{0}
\renewcommand{\thefigure}{S\arabic{figure}}

\setcounter{table}{0}
\renewcommand{\thetable}{S\arabic{table}}


\clearpage
\onecolumngrid
\section{Supplementary Material}

This supplementary material provides methodological details and technical specifications to ensure reproducibility of the results presented in the main manuscript. First, we present a detailed derivation of the pion-nucleon sigma term ($\sigma_{\pi N}$) in baryon chiral perturbation theory (BChPT). Next, we systematically analyze the finite volume correction (FVC) required for lattice quantum chromodynamics (QCD) data. Finally, more fitting results at different chiral orders are provided for comparison.

\subsection{Calculation of pion-nucleon sigma term in BChPT}

The pion-nucleon sigma term $\sigma_{\pi N}$ is defined by the scalar form factor $\sigma(t)$ of the nucleon at zero momentum transfer,
\begin{align}
    \sigma_{\pi N}=\sigma(t=0)=\frac{1}{2 m_N}\langle N(p^\prime)|\hat{m}(\bar{u}u+\bar{d}d)|N(p)\rangle|_{p=p^\prime}\ ,\quad t=(p^\prime-p)^2\ ,\label{eq.sigma.def}
\end{align}
where the normalization of the nucleon spinors is set to $\bar{u}(p,s^\prime)u(p,s)=2m_N\delta_{s^\prime s}$, and $\hat{m}=(m_u+m_d)/2$ represents the average of the $u$ and $d$ quark masses. The above definition of $\sigma_{\pi N}$ is scale-independent and serves as a measure of explicit chiral symmetry breaking. On the other hand, it follows from the Feynman-Hellmann (FH) theorem~\cite{Hellmann:1937book,Feynman:1939zza} that
\begin{align}
    \frac{\partial m_N^2}{\partial \hat{m}} =
    \langle N(p)|\bar{u}u+\bar{d}d|N(p)\rangle\ .\label{eq.FH.original}
\end{align}
In view of Eqs.~\eqref{eq.sigma.def} and~\eqref{eq.FH.original}, $\sigma_{\pi N}$ can also be obtained from the derivative of the nucleon mass,
\begin{align}
    \sigma_{\pi N} = \hat{m}\frac{\partial m_N}{\partial \hat{m}}  = M^2\frac{\partial m_N}{\partial M^2}\ ,\label{eq.FH.mn}
\end{align}
where $M^2=2 B_0 \hat{m}$ is the leading-order (LO) pion mass with $B_0$ being a constant associated with the quark condensate. Therefore, there are two approaches to obtain the nucleon sigma term: either by direct computation of the scalar form factor at $t=0$ as specified by Eq.~\eqref{eq.sigma.def} (direct approach) or via the derivative of the nucleon mass given by Eq.~\eqref{eq.FH.mn} (FH approach). Up to and including $\mathcal{O}(p^5)$ in BChPT, they share the same set of effective operators that can be found in Ref.~\cite{Liang:2025cjd}. 

\begin{figure}[tbh]
\centering
\includegraphics[width=0.85\textwidth, angle=-0]{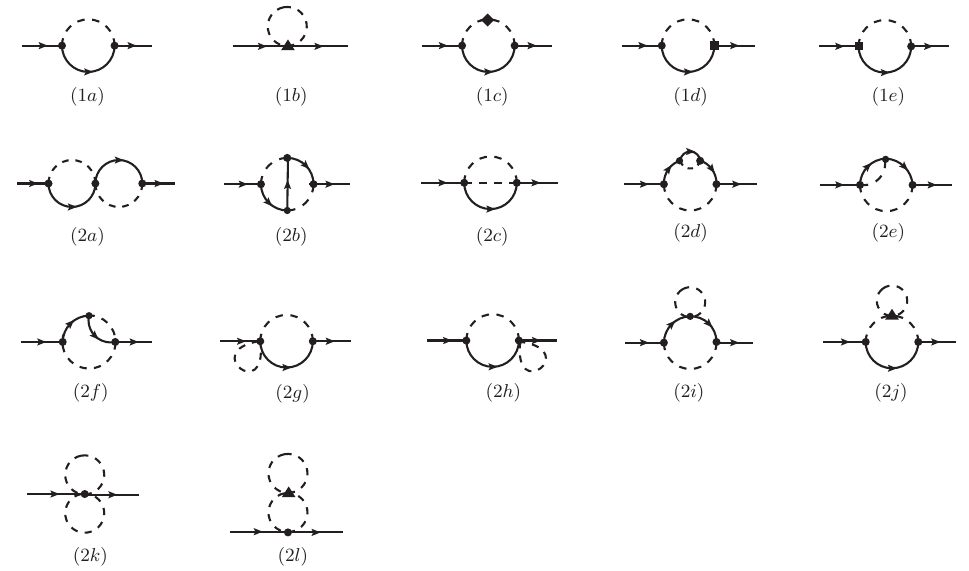}
\caption{One-loop and two-loop Feynman diagrams contributing to the nucleon mass and $\pi N$ sigma term $\sigma_{\pi N}$. The solid and dashed lines denote nucleons and pions, respectively.
The dots, triangle, square and diamond represent the vertices of $\mathcal{O}(p^1)$, $\mathcal{O}(p^2)$, $\mathcal{O}(p^3)$ and $\mathcal{O}(p^4)$, in order.}
\label{fig.loop}
\end{figure}

Below, we describe the calculation of $\sigma_{\pi N}$ in BChPT up to leading two-loop order using these two approaches. 
\begin{itemize}
    \item FH approach: The one- and two-loop Feynman diagrams relevant to the nucleon self-energy are displayed in Fig.~\ref{fig.loop} (Fig.~\ref{fig.loops_onecol} in the main text). From these diagrams, contributions to the nucleon mass can be derived, as done very recently in Ref.~\cite{Liang:2025cjd}. Furthermore, in the FH approach, taking the derivative with respect to the LO pion mass yields the chiral expressions of the pion-nucleon sigma term.
    \item Direct approach: The Feynman diagrams required for deriving $\sigma_{\pi N}$ in the direct approach can be obtained by attaching an isoscalar-scalar external current to the self-energy diagrams of Fig.~\ref{fig.loop} in all possible ways. In addition, the wave-function renormalization constant must be taken into account to incorporate chiral corrections on the external nucleon legs. The same procedure has also been applied to the calculation of the axial-vector coupling constant $g_A$ at two-loop order in Ref.~\cite{Bernard:2025gto}.
\end{itemize}

Both approaches lead to the same result for $\sigma_{\pi N}$:
\begin{align}\label{eq.simga.term.full.unren}
    \sigma_{\pi N} = -4{c}_1 M^2-4{e}_m M^4 + \Delta \sigma_{\pi N}^{(1)}+
   \Delta \sigma_{\pi N}^{(2)}\ ,
\end{align}
where $\Delta \sigma_{\pi N}^{(1)}$ and $\Delta \sigma_{\pi N}^{(2)}$ denote the contributions from one- and two-loop Feynman diagrams, respectively. The one-loop contribution $\Delta \sigma_{\pi N}^{(1)}$ can be expressed as
\begin{align}\label{eq.sigma.term.olo}
    \Delta \sigma_{\pi N}^{(1)} = 
    \sigma_{\pi N}^{(1a)}
    +\sigma_{\pi N}^{(1b)}
    +\sigma_{\pi N}^{(1c)}
    +\sigma_{\pi N}^{(1d)}
    +\sigma_{\pi N}^{(1e)} \ , 
\end{align}
where
\begin{align}
\sigma_{\pi N}^{(1a)}  = &-\frac{3i g_A^2 M^2 m (J_{11}+M^2 J_{21})}{2F^2} \ ,\label{eq.sigma.term.1a}\\
\sigma_{\pi N}^{(1b)}  =& -\frac{3i\left[c_2+(-2 c_1+c_3)d\right] M^2(J_{10}+M^2J_{20})}{d F^2} \ ,\\
\sigma_{\pi N}^{(1c)}  = &-\frac{3 i g_A^2 M^2 m}{2 F^4 \left(M^2-4 m^2\right)^2} \left\{2 J_{01} \left((d-2) l_3 M^2 (M^2-8 m^2) -l_4 (M^2-4 m^2)^2\right) \right. \notag\\  
& \left.+M^2 \left[-(d-2) l_3 (M^2-8 m^2) J_{10}+4 J_{11} \left(l_3 \left((13-7 d) M^2 m^2+(d-2) M^4\right.\right.\right.\right. \notag \\ 
& \left.\left.\left.\left.+8 (d-1) m^4\right)-l_4 \left(M^2-4 m^2\right)^2\right)+M^2 \left(M^2-4 m^2\right) \left(2 J_{21} \left((d-2) l_3 M^2 \right.\right.\right.\right. \notag \\ & \left.\left.\left.\left.-2 (d-1) l_3 m^2 -l_4 \left(M^2-4 m^2\right)\right)-(d-2) l_3 J_{20}\right)\right]\right\} \ ,\\
\sigma_{\pi N}^{(1d)}  = & \sigma_{\pi N}^{(1e)}=-\frac{3 i g_A M^2 m (2 d_{16}-d_{18})}{F^2}\left(M^4 J_{21}+2 M^2 J_{11}+J_{01}\right) \, ,
\end{align}
where $m$, $g_A$ and $F$ represent the nucleon mass, nucleon axial-vector coupling constant and pion decay constant in the chiral limit, respectively, the parameters $c_{1,2,3}$ and $d_{16,18}$ are low-energy constants (LECs) stemming from the $\mathcal{O}(p^2)$ and $\mathcal{O}(p^3)$ pion-nucleon Lagrangians~\cite{Yao:2016vbz}, and $l_{3,4}$ are $\mathcal{O}(p^4)$ mesonic LECs~\cite{Gasser:1983yg}.
The one-loop integrals are defined by
\begin{align}
    J_{\nu_1\nu_2}=\kappa\int \frac{d^d \ell}{ i\pi^{d/2}} \frac{1}{[\ell^2 - M^2 + i0^+]^{\nu_1} [(\ell + p)^2 - m^2 + i0^+]^{\nu_2}} \ ,\quad p^2= m^2\ ,
\end{align}
where $\kappa=i\mu^{2\epsilon}\exp[\epsilon(\gamma_E-1)]/\Lambda$ with $\epsilon=(4-d)/2$ and $d$ the space-time dimension. Furthermore, $\mu$ and $\gamma_E$ are the renormalization scale in dimensional regularization and the Euler constant, respectively. The sum of the two-loop contributions $\Delta \sigma_{\pi N}^{(2)}$ is given by
\begin{align}\label{eq.sigma.term.tlo}
    \Delta \sigma_{\pi N}^{(2)} = 
    \sigma_{\pi N}^{(2a)}
    +\sigma_{\pi N}^{(2b)}
    +\cdots
    +\sigma_{\pi N}^{(2l)} 
    +\sigma_{\pi N}^{(2\prime)}
    +\sigma_{\pi N}^{\rm sub.}\ . 
\end{align}
Here $\sigma_{\pi N}^{(2\prime)}=M^2 \partial(m_N^{(1a)}\delta \mathcal{Z}_N^{(2)})/\partial M^2$, where $m_N^{(1a)}$ and $\delta \mathcal{Z}_N^{(2)}$ are given in Ref.~\cite{Liang:2025cjd}. The last term $\sigma_{\pi N}^{\rm sub.}$ incorporates finite pieces generated by one-loop diagrams that appear as sub-diagrams within two-loop diagrams; see the discussions around Eq.~(3.18) of Ref.~\cite{Liang:2025cjd}. The remaining terms in Eq.~\eqref{eq.sigma.term.tlo} are listed below:
\begin{align}
\sigma_{\pi N}^{(2a)}  &= \frac{3 g_A^2 M^2 m}{2 F^4}( M^4 I_{12011}+M^4 I_{21011}+M^2 I_{02011}+2 M^2 I_{11011}+M^2 I_{20011}+I_{10011}+I_{01011})\ ,\\
\sigma_{\pi N}^{(2b)}  &=\frac{g_A^4 M^2}{32 F^4 m(d-2)(3 d-4)}\left\{ -(d-2)^2 I_{01100} -48(d-2)(3 d-4) m^4 I_{01111}-12 (d-2)(3 d-4)m^2 I_{02010}\right.\notag\\ \notag 
&\left.+\left(4 (d-2)(2 d-3) m^2-(d-2)^2 M^2\right)I_{02100} -48(d-2)(3 d-4) M^2 m^4 I_{02111} -(d-2)^2 I_{10100}\right.\\ \notag &\left.-48(d-2)(3 d-4) m^4 I_{10111} +2(d-2)^2 I_{11000}-12(d-2)(3 d-4) m^2 I_{11001}-12(d-2)(3 d-4) m^2 I_{11010}  \right.\\ \notag &\left.+\left(4 \left(2 \left(d^2-5 d+6\right) M^2+\left(-19 d^2+75 d-64\right) m^2\right)\right)I_{11100}-96(d-2)(3 d-4) M^2 m^4 I_{11111}\right.\\ \notag & \left.-8 (d-2)M^2 \left(M^2-m^2\right)I_{12100}  -\left(8 M^2  \left((d-2) M^2+2 (9-7 d) m^2\right)\right)I_{12100}-\left(8 \left(M^2-m^2\right)  \left((d-2) M^2\right.\right.\right.\\ \notag &\left.\left.\left.+2 (9-7 d) m^2\right)\right)I_{12100}  -12(d-2)(3 d-4) m^2 I_{20001}+ \left(4 (d-2)(2 d-3) m^2-(d-2)^2 M^2\right)I_{20100}\right.\\ \notag &\left.-48 (d-2)(3 d-4)M^2 m^4 I_{20111}+2  \left((d-2)^2 M^2+(7d-9)(d-2) m^2\right)(I_{12000}+I_{21000})-12(d-2)\right.\\ \notag &\left.(3 d-4) M^2 m^2 (I_{12001}+I_{21001})-12 (d-2)(3 d-4)M^2 m^2 (I_{12010}+I_{21010})-8 (d-2)M^2 \left(M^2-m^2\right)\right.\\ \notag &\left.I_{21100}-8 M^2  \left((d-2) M^2+2 (9-7 d) m^2\right)I_{21100}-8 \left(M^2-m^2\right)  \left((d-2) M^2+2 (9-7 d) m^2\right)I_{21100}\right.\\ \notag &\left.+4  \left(\left(-19 d^2+75 d-64\right) M^2 m^2+\left(d^2-5 d+6\right) M^4+2 \left(14 d^2-53 d+45\right) m^4\right) (I_{12100}+I_{21100})\right.\\ \notag &\left.-48(d-2)(3 d-4) M^4 m^4 (I_{12111}+I_{21111})-8 M^2 \left(M^2-m^2\right)  \left((d-2) M^2 +2 (9-7 d) m^2\right)\right.\\  &\left.(2 I_{13100}+I_{22100})-8 M^2 \left(M^2-m^2\right)\left((d-2) M^2+2 (9-7 d) m^2\right) (I_{22100}+2 I_{31100})\right\}  \ ,\\
\sigma_{\pi N}^{(2c)}  &= -\frac{1}{8 (d-2) (3 d-4) F^4 m}\left\{M^2 \left[ (d-2)^2 I_{01100}+(d-2) I_{02100} \left((d-2) M^2+4 (d-1) m^2\right)\right.\right.\notag\\ \notag &\left.\left.+(d-2)^2 I_{10100}-2 (d-2)^2 I_{11000}-4  \left(2 \left(d^2-5 d+6\right) M^2+\left(-7 d^2+23 d-16\right) m^2\right)I_{11100}\right.\right.\\ \notag &\left.\left.+(d-2)  \left((d-2) M^2+4 (d-1) m^2\right)I_{20100}-2 (d-2)\left((d-2) M^2+(d-1) m^2\right)(I_{12000}+I_{21000})\right.\right.\\ \notag &\left.\left.+8 (d-2) M^2 \left(M^2-m^2\right) (I_{12100}+I_{21100})+8 M^2  \left((d-2) M^2-2 (d-1) m^2\right)(I_{12100}+I_{21100})\right.\right.\\ \notag &\left.\left.+8 \left(M^2-m^2\right)  \left((d-2) M^2-2 (d-1) m^2\right)(I_{12100}+I_{21100})-4  \left(\left(-7 d^2+23 d-16\right) M^2 m^2\right.\right.\right.\\ \notag &\left.\left.\left.+\left(d^2-5 d+6\right) M^4+2 \left(2 d^2-7 d+5\right) m^4\right)(I_{12100}+I_{21100})+16 M^2 \left(M^2-m^2\right)  \left((d-2) M^2\right.\right.\right.\\  &\left.\left.\left.-2 (d-1) m^2\right)(I_{13100}+I_{22100}+I_{31100})\right] \right\}\ ,\\
\sigma_{\pi N}^{(2d)} & = \frac{1}{32 (d-2) (3 d-4) F^4 m \left(M^2-4 m^2\right)^2} \left\{ 3 g_A^4 M^2\left\{-24 (4-3 d) (2-d) m^2  \left((d-3) M^2+4 (d-1) m^2\right)\right.\right.\notag \\
&\left.\left.I_{00110}-12 (4-3 d) (2-d) m^2 \left((3-2 d) M^2+4 (d-1) m^2\right)I_{01010}-(2-d)  \left(16 \left(9 d^2-22 d+13\right) m^4\right.\right.\right.\notag\\  
&\left.\left.\left.+4 (13-9 d) M^2 m^2+(d-2) M^4\right)I_{01100}-24 (4-3 d) (2-d) M^2 m^2 \left((d-3) M^2+2 (d-1) m^2\right)I_{01110}\right.\right.\notag\\  &\left.\left.-(2-d)  \left(16 \left(9 d^2-22 d+13\right) m^4+4 (13-9 d) M^2 m^2+(d-2) M^4\right)I_{10100}-24 (4-3 d) (2-d) M^2 m^2 \right.\right.\notag\\ 
&\left.\left.\left((d-3) M^2+2 (d-1) m^2\right)I_{10110}-2 (d-2) \left(M^2-4 m^2\right) \left((d-2) M^2+(d-1) m^2\right)I_{11000}-12 (4-3 d) \right.\right.\notag\\  
&\left.\left.(2-d) M^2 m^2 \left((3-2 d) M^2+4 (d-1) m^2\right)I_{11010}+4 \left(\left(54 d^3-331 d^2+641 d-392\right) M^4 m^2+2 \left(20 d^2\right.\right.\right.\right.\notag\\  &\left.\left.\left.\left.-67 d+59\right) M^2 m^4-\left(d^2-5 d+6\right) M^6+8 \left(2 d^2-7 d+5\right) m^6\right)I_{11100} -24 (d-3) (d-2) (3 d-4) M^6\right.\right.\notag\\  
&\left.\left. m^2 I_{11110}-8 M^2 \left(m^2-M^2\right) \left(-2 \left(18 d^2-57 d+43\right) M^2 m^2+(d-2) M^4+8 (d-1) m^4\right) I_{12100}\right.\right.\notag\\  &\left.\left.-8 M^2 \left(m^2-M^2\right) \left(-2 \left(18 d^2-57 d+43\right) M^2 m^2+(d-2) M^4+8 (d-1) m^4\right)I_{21100}-\left(4 m^2-M^2\right)\right.\right.\notag\\  
&\left.\left.\left[24 (d-3) (d-2) (3 d-4) m^2 I_{00110}-12 (d-2) (2 d-3) (3 d-4) m^2 I_{01010}+(2-d) \left(2 (d-2) M^2\right.\right.\right.\right.\notag\\  &\left.\left.\left.\left.+4 (13-9 d) m^2\right)I_{01100}+24 (d-3) (d-2) (3 d-4) M^2 m^2 I_{01110} +24 (4-3 d) (2-d) m^2 \left((d-3) M^2\right.\right.\right.\right.\notag\\  &\left.\left.\left.\left.+2 (d-1) m^2\right) I_{01110} +(2-d) \left(2 (d-2) M^2+4 (13-9 d) m^2\right)I_{10100}+24 (d-3) (d-2) (3 d-4) M^2 m^2 \right.\right.\right.\notag\\  &\left.\left.\left.I_{10110}+24 (4-3 d) (2-d) m^2 \left((d-3) M^2+2 (d-1) m^2\right)I_{10110} +2 (d-2)^2 \left(M^2-4 m^2\right) I_{11000} +2 (d-2) \right.\right.\right.\notag\\  &\left.\left.\left.\left((d-2) M^2+(d-1) m^2\right)I_{11000}-12 (d-2) (2 d-3) (3 d-4) M^2 m^2 I_{11010}+12 (4-3 d) (2-d) m^2  \right.\right.\right.\notag\\  &\left.\left.\left.\left((3-2 d) M^2+4 (d-1) m^2\right)I_{11010}-4 \left(2 \left(54 d^3-331 d^2+641 d-392\right) M^2 m^2-3 \left(d^2-5 d+6\right) M^4\right.\right.\right.\right.\notag\\  &\left.\left.\left.\left.+2 \left(20 d^2-67 d+59\right) m^4\right)I_{11100}+72 (d-3) (d-2) (3 d-4) M^4 m^2 I_{11110}+8M^2 \left(m^2-M^2\right) \left(2 (d-2) M^2\right.\right.\right.\right.\notag\\  &\left.\left.\left.\left.-2 \left(18 d^2-57 d+43\right) m^2\right)I_{12100}-8M^2  \left(-2 \left(18 d^2-57 d+43\right) M^2 m^2+(d-2) M^4+8 (d-1) m^4\right)I_{12100}\right.\right.\right.\notag\\  &\left.\left.\left.+8 \left(m^2-M^2\right) \left(-2 \left(18 d^2-57 d+43\right) M^2 m^2+(d-2) M^4+8 (d-1) m^4\right)I_{12100}+8 M^2 \left(m^2-M^2\right)\right.\right.\right.\notag\\  &\left.\left.\left. \left(2 (d-2) M^2-2 \left(18 d^2-57 d+43\right) m^2\right)I_{21100}-8 M^2 \left(-2 \left(18 d^2-57 d+43\right) M^2 m^2+(d-2) M^4\right.\right.\right.\right.\notag\\  &\left.\left.\left.\left.+8 (d-1) m^4\right)I_{21100}+8 \left(m^2-M^2\right)  \left(-2 \left(18 d^2-57 d+43\right) M^2 m^2+(d-2) M^4+8 (d-1) m^4\right)I_{21100}\right]\right.\right.\notag\\  &\left.\left.-\left(4 m^2-M^2\right)\left[ 12 (4-3 d) (2-d) m^2 \left((3-2 d) M^2+4 (d-1) m^2\right)I_{02010} +(2-d) \left(16 \left(9 d^2-22 d+13\right) m^4\right.\right.\right.\right.\notag\\  &\left.\left.\left.\left.+4 (13-9 d) M^2 m^2+(d-2) M^4\right)I_{02100}+24 (4-3 d) (2-d) M^2 m^2 \left((d-3) M^2+2 (d-1) m^2\right)I_{02110}\right.\right.\right.\notag\\  &\left.\left.\left.+(2-d) \left(16 \left(9 d^2-22 d+13\right) m^4+4 (13-9 d) M^2 m^2+(d-2) M^4\right)I_{20100}+24 (4-3 d) (2-d) M^2 m^2 \right.\right.\right.\notag\\  &\left.\left.\left.\left((d-3) M^2+2 (d-1) m^2\right)I_{20110}+2 (d-2) \left(M^2-4 m^2\right) \left((d-2) M^2+(d-1) m^2\right)(I_{12000}+I_{21000})\right.\right.\right.\notag\\  &\left.\left.\left.+12 (4-3 d) (2-d) M^2 m^2 \left((3-2 d) M^2+4 (d-1) m^2\right)(I_{12010}+I_{21010})-4 \left(\left(54 d^3-331 d^2+641 d-392\right) \right.\right.\right.\right.\notag\\  &\left.\left.\left.\left.M^4 m^2+2 \left(20 d^2-67 d+59\right) M^2 m^4-\left(d^2-5 d+6\right) M^6+8 \left(2 d^2-7 d+5\right) m^6\right) (I_{12100}+I_{21100})\right.\right.\right.\notag\\  &\left.\left.\left.+24 (d-3) (d-2) (3 d-4) M^6 m^2 (I_{12110}+I_{21110})+8 M^2 \left(m^2-M^2\right) \left(-2 \left(18 d^2-57 d+43\right) M^2 m^2\right.\right.\right.\right.\notag\\  &\left.\left.\left.\left.+(d-2) M^4+8 (d-1) m^4\right)(2 I_{13100}+I_{22100})+8 M^2 \left(m^2-M^2\right) \left(-2 \left(18 d^2-57 d+43\right) M^2 m^2\right.\right.\right.\right.\notag\\  &\left.\left.\left.\left.+(d-2) M^4+8 (d-1) m^4\right)(I_{22100}+2 I_{31100})\right]\right\} \right\} \ , \\
\sigma_{\pi N}^{(2e)}  &=\sigma_{\pi N}^{(2f)} = \frac{1}{8 (d-2) (3 d-4) F^4 m}\left\{g_A^2 M^2 \left\{(d-2)^2 I_{01100}+12 (d-2) (3 d-4) m^2 I_{01101}-(2-d)  \left((d-2) M^2\right.\right.\right.\notag \\ &\left.\left.\left.+4 (3-2 d) m^2\right)I_{02100}+12 (d-2) (3 d-4) M^2 m^2 I_{02101}+(d-2)^2 I_{10100}+12 (d-2) (3 d-4) m^2 I_{10101}\right.\right. \notag\\ &\left.\left.-2 (d-2)^2 I_{11000}-4 \left(2 \left(d^2-5 d+6\right) M^2+\left(8 d^2-33 d+32\right) m^2\right)I_{11100}+24 (d-2) (3 d-4) M^2 m^2 I_{11101}\right.\right.\notag\\ 
&\left.\left.+8 (d-2) M^2 \left(M^2-m^2\right) I_{12100}+8 M^2 \left((d-2) M^2+(4 d-6) m^2\right)I_{12100}+8 \left(M^2-m^2\right) \left((d-2) M^2\right.\right.\right.\notag\\  
&\left.\left.\left.+(4 d-6) m^2\right)I_{12100}-(2-d) \left((d-2) M^2+4 (3-2 d) m^2\right)I_{20100}+12 (d-2) (3 d-4) M^2 m^2 I_{20101}\right.\right.\notag\\  &\left.\left.+2 (2-d) \left((d-2) M^2+(3-2 d) m^2\right)(I_{12000}+I_{21000})+8 (d-2) M^2 \left(M^2-m^2\right) I_{21100}+8 M^2 \left((d-2) M^2\right.\right.\right.\notag\\ 
&\left.\left.\left.+(4 d-6) m^2\right)I_{21100}+8 \left(M^2-m^2\right) \left((d-2) M^2+(4 d-6) m^2\right)I_{21100}-4 \left(\left(8 d^2-33 d+32\right) M^2 m^2\right.\right.\right.\notag\\  &\left.\left.\left.+\left(d^2-5 d+6\right) M^4-2 \left(4 d^2-16 d+15\right) m^4\right)(I_{12100}+I_{21100})+12 (d-2) (3 d-4) M^4 m^2 (I_{12101}+I_{21101})\right.\right.\notag \\
&\left.\left.+8 M^2 \left(M^2-m^2\right) \left((d-2) M^2+(4 d-6) m^2\right) (2 I_{13100}+I_{22100})+8 M^2 \left(M^2-m^2\right) \left((d-2) M^2\right.\right.\right.\notag\\ 
&\left.\left.\left.+(4 d-6) m^2\right)(I_{22100}+2 I_{31100})\right\}\right\}\ , \\
\sigma_{\pi N}^{(2g)}  & =\sigma_{\pi N}^{(2h)}=-\frac{1}{2F^4}g_A^2 M^2 m \left(M^2 I_{12001}+M^2 I_{21001}+I_{11001}+I_{20001}\right)\ ,\\
\sigma_{\pi N}^{(2i)}  &=0 \  ,\\
\sigma_{\pi N}^{(2j)}  & =\frac{1}{8 F^4 \left(M^2-4 m^2\right)^2}g_A^2 M^2 m \left\{  \left((4-6 d) M^2+32 m^2\right)I_{10001}-\left(M^2-4 m^2\right) \left[4\left((2-3 d) M^2+(3 d+5) m^2\right)\right.\right.\notag\\ 
 &\left.\left.I_{11001}+(4-6 d) I_{10001}+3 (d-2) I_{11000}\right]+M^2 \left[2 \left((2-3 d) M^2+2 (3 d+5) m^2\right)I_{11001}+3 (d-2) I_{11000}\right]\right.\notag \\ 
&\left.-\left(M^2-4 m^2\right) \left[M^2 \left(2 \left((2-3 d) M^2+2 (3 d+5) m^2\right)(I_{12001}+I_{21001})+3 (d-2) (I_{12000}+I_{21000})\right)\right.\right. \notag\\
&\left.\left.+ \left((4-6 d) M^2+32 m^2\right)I_{20001}\right]\right\} \  ,\\
\sigma_{\pi N}^{(2k)}  &=0 \  ,\\
\sigma_{\pi N}^{(2l)}  &=0 \  .
\end{align}
The two-loop diagrams involve at most five independent internal propagators, and hence the two-loop integrals are generically defined as
\begin{align}
I_{\nu_1\nu_2\nu_3\nu_4\nu_5} =\kappa^2 \int\frac{{\rm d}\ell_1^d}{{ i}\pi^{d/2}}\frac{{\rm d}\ell_2^d}{{i}\pi^{d/2}}\frac{1}{\mathcal{D}_1^{\nu_1}}\frac{1}{\mathcal{D}_2^{\nu_2}}\frac{1}{\mathcal{D}_3^{\nu_3}}\frac{1}{\mathcal{D}_4^{\nu_4}}\frac{1}{\mathcal{D}_5^{\nu_5}}\ , \quad p^2=m^2\ ,
\end{align}
where $\nu_i$ are integers and the denominators are given by
\begin{align}
 &\mathcal{D}_1=\ell_1^2-M^2+{i}0^+ ,\;\;  \mathcal{D}_2=\ell_2^2-M^2+{i}0^+ , \;\;
 \mathcal{D}_3=(\ell_1+\ell_2+p)^2-m^2+{ i}0^+ ,\;\;\notag\\
 &\mathcal{D}_4=(\ell_1+p)^2-m^2+i 0^+ ,\;\;
\mathcal{D}_5=(\ell_2+p)^2-m^2+{i}0^+ .
\end{align}
All loop integrals can be reduced to a set of master integrals, consisting of 13 independent ones, by using the method of integration by parts (IBP) (see Ref.~\cite{Liang:2025cjd} and the references therein for more details).

The renormalization of $\sigma_{\pi N}$ at two-loop order is carried out using dimensional regularization (DR) and the extended-on-mass-shell (EOMS) scheme~\cite{Fuchs:2003qc}. The ultraviolet (UV) divergences are handled within DR, while the power counting breaking (PCB) terms are removed using the EOMS scheme. The renormalization procedure for $\sigma_{\pi N}$ is the same as that for the nucleon mass, which has been detailed in Ref.~\cite{Liang:2025cjd}.

In Eq.~\eqref{eq.simga.term.full.unren}, $M^2$ is the LO pion mass. 
However, the pion mass computed in lattice QCD (LQCD) is the full pion mass, denoted as $M_\pi$. It is usually calculated up to a specific chiral order in chiral perturbation theory (ChPT) by systematically incorporating corrections such as pion loops and counterterms. For our current accuracy, the $\mathcal{O}(p^4)$ chiral expression of $M_\pi$ is required, which reads~\cite{Gasser:1983yg} 
\begin{align}
  M^2  = M_\pi^2\bigg[1-\Delta_{M_\pi}^{(2)}\bigg]\ ,\quad \Delta_{M_\pi}^{(2)}\equiv\frac{2l_3^r M_\pi^2}{F^2}+\frac{M_\pi^2}{F^2\Lambda}\log\frac{M_\pi}{\mu}\, ,\label{eq.pion.mass.p4}
\end{align}
where $l_3^r=(1.21\pm 0.01)\times 10^{-3}$ with the renormalization scale $\mu$ taken at the physical nucleon mass~\cite{FlavourLatticeAveragingGroupFLAG:2024oxs}. 
The superscript of $\Delta_{M_\pi}^{(2)}$ denotes the chiral dimension of this piece. Expanding $\sigma_{\pi N}(M^2)$ around $M_\pi^2$, we have
\begin{align}
    \sigma_{\pi N}(M^2) = \sigma_{\pi N}(M_\pi^2)+\dot{\sigma}_{\pi N}(M_\pi^2)(M^2-M_{\pi}^2)+\cdots
\end{align}
where $\dot{\sigma}_{\pi N}(M_\pi^2)\equiv \big[{\partial \sigma_{\pi N}}/{\partial M^2}\big]_{M^2=M^2_\pi}$, being the derivative of ${\sigma}_{\pi N}$ with respect to $M^2$,  evaluated at $M^2=M_\pi^2$. Inserting Eq.~\eqref{eq.pion.mass.p4} into the above equation and performing a chiral expansion, we obtain
\begin{align}
    \sigma_{\pi N}(M^2) = \sigma_{\pi N}(M_\pi^2)- \underbrace{\dot{\sigma}_{\pi N} ^{(2)}(M_\pi^2)\,M_{\pi}^2\,\Delta_{M_\pi}^{(2)}}_{\mathcal{O}(p^4)}- \underbrace{\dot{\sigma}_{\pi N} ^{(1a)}(M_\pi^2)\,M_{\pi}^2\,\Delta_{M_\pi}^{(2)}}_{\mathcal{O}(p^5)}+\text{higher-order pieces beyond $\mathcal{O}(p^5)$}\ ,\label{eq.sigma.term.physical.pi}
\end{align}
where $\dot{\sigma}_{\pi N} ^{(2)}(M_\pi^2) =  - 4\tilde{c}_1$, and $\dot{\sigma}_{\pi N} ^{(1a)}(M_\pi^2)$ is the slope of the sigma term ${\sigma}_{\pi N} ^{(1a)}$ in Eq.~\eqref{eq.sigma.term.1a}. 

\section{Estimate of finite volume correction}

Finite volume correction is one of the systematic uncertainties in predictions from lattice QCD simulations. A standard approach to estimate FVC is to utilize a version of ChPT generalized from infinite space to a finite spatial box. The FVC to the nucleon mass has been calculated at one-loop level in a series of works~\cite{Beane:2004tw,Geng:2011wq,Ren:2012aj,Alvarez-Ruso:2013fza,Liang:2022tcj}, from which the FVC to the pion-nucleon sigma term can be derived via the FH theorem.

Here, we employ the result provided in Ref.~\cite{Liang:2022tcj}, where the FVC is calculated in covariant BChPT up to $\mathcal{O}(p^3)$ with explicit $\Delta(1232)$ resonances, which partially simulate two-loop contributions since no two-loop FVC calculations are currently available for the problem at hand.\footnote{Strict ChPT calculations of FVC at two-loop level have only been realized for the meson sector~\cite{Colangelo:2006mp,Bijnens:2014dea}, and the results indicate that the two-loop contribution is suppressed compared to the leading-loop one.} The result is expressed in a very compact form due to the introduction of a unified formulation for one-loop integrals in the presence of finite volume effects. Specifically, the FVC to the nucleon mass stemming from the $\pi N$ loop reads
\begin{align}
\Delta_L m_N^{(N)} (M_{\pi}; L)=\frac{3g_A^2 m}{2F^2}\big[\widetilde{A}_0(m^2;L)+M_{\pi}^2 \widetilde{B}_0(m^2,m^2,M_{\pi}^2;L)\big] \ ,\label{eq.olo.mn.piN.FVC}
\end{align}
while contribution from the $\pi\Delta$ loop is
\begin{align}
\hspace{-0.2cm}
\Delta_L m_N^{(\Delta)} (M_{\pi}; L)&=\frac{h_A^2}{6F^2 m^2_\Delta m} \bigg\{ 
\big[m_\Delta^4+2m_\Delta^3 m - 2m_\Delta^2(m^2+M_{\pi}^2)+2m_\Delta m(m^2-M_{\pi}^2)+(m^2-M_{\pi}^2)^2 \big] \widetilde{A}_0(m_\Delta^2 ;L)
\notag\\
&+\big[2M_{\pi}^2(m_\Delta^2-3m_\Delta m-2m^2)-(m_\Delta-m)(m_\Delta+m)^3-M_{\pi}^4 \big]\widetilde{A}_0(M_{\pi}^2;L)
+4m^2 \widetilde{A}_{00}(m_\Delta^2;L)
\notag \\
&-4m^2 \widetilde{A}_{00}(M_{\pi}^2;L)
-\big((m_\Delta-m)^2-M_{\pi}^2\big)(m_\Delta+m-M_{\pi})^2(m_\Delta+m+M_{\pi})^2\widetilde{B}_0(m^2, m_\Delta^2,M_{\pi}^2;L)
\bigg\}
\ .
\label{eq.olo.mn.piD.FVC}
\end{align}
Explicit expressions of the loop integrals, $\widetilde{A}_{0}$, $\widetilde{A}_{00}$ and $\widetilde{B}_{0}$, can be found in the appendix of Ref.~\cite{Liang:2022tcj}. 

Using Eq.~\eqref{eq.olo.mn.piN.FVC}, the FVC to $\sigma_{\pi N}$ can be obtained straightforwardly by applying the FH theorem, i.e.,
\begin{align}
\Delta_L \sigma_{\pi N}^{(N)} (M_{\pi};L) = \sum _{n_s}\vartheta(n_s)\int_0^1 {\rm d}x_1\frac{3g_A^2 m M_{\pi}^2}{32 F^2\pi^2}\bigg[2K_0(\sqrt{\mathcal{M}^2_N})+\frac{L^2 M_{\pi}^2 n_s(x_1-1)}{\sqrt{\mathcal{M}^2_N}}K_1(\sqrt{\mathcal{M}^2_N})\bigg]\ ,
\end{align}
where $\mathcal{M}_N^2=L^2n_s[(1-x_1)M_{\pi}^2+x_1^2 m^2]$ and $\vartheta(n_s)$ denotes the multiplicity of the vector $\mathbf{n}=(n_x,n_y,n_z)$ for $n_s\equiv n_x^2+n_y^2+n_z^2$ with integers $n_{x,y,z}\in \mathbb{Z}$. $K_z$ is the modified Bessel function of the second kind. A simultaneous expansion in large $L$ and large $m$ yields the following asymptotic formula~\cite{Beane:2004tw},
\begin{align}
 \Delta_L \sigma_{\pi N}^{(N)} (M_{\pi};L) \underset{L\to \infty}{\longrightarrow}   b_L\left(\frac{M_{\pi}^3}{M_{\pi} L}-\frac{M_{\pi}^3}{2}\right) \exp(-M_{\pi} L) \ , \quad b_L={9g_A^2}/{(8\pi F^2)}\sim 68~\textrm{GeV}^{-2}\ .
\end{align}
This equation has been employed in Ref.~\cite{Agadjanov:2023efe} to extrapolate the finite-volume results to the infinite-volume limit, treating $b_L$ as a free parameter.

Likewise, for the $\Delta$ case, with Eq.~\eqref{eq.olo.mn.piD.FVC} one has
\begin{align}
\Delta_L \sigma_{\pi N}^{(\Delta)}(M_{\pi};L)&=\sum _{n_s}\vartheta(n_s)\int_0^1 {\rm d}x_1
\frac{h_A^2 M_{\pi}^2}{96\pi^2 F^2 m_\Delta^2 m}  
\bigg\{ 
2\big((m_\Delta+m)^2-M_{\pi}^2\big)^2 K_0(\sqrt{\mathcal{M}_\Delta^2})
\notag\\
&
+4\big((m_\Delta+m)^2-M_{\pi}^2\big)\big((m_\Delta-m)^2-M_{\pi}^2\big)K_0(\sqrt{\mathcal{M}_\Delta^2})
\notag \\
&-\frac{L^2 n_s(x_1-1)\big((m_\Delta+m)^2-M_{\pi}^2\big)^2\big((m_\Delta-m)^2-M_{\pi}^2\big)}{\sqrt{\mathcal{M}^2_\Delta}} K_1(\sqrt{\mathcal{M}_\Delta^2})
\notag\\
&
+\frac{8m_\Delta \big(m_\Delta^2+m_\Delta m+m^2-M_{\pi}^2\big)}{L\sqrt{n_s}}K_1(m_\Delta L \sqrt{n_s})
+\frac{8M_{\pi}(M_{\pi}^2+2m^2-m_\Delta^2+3m_\Delta m)}{L\sqrt{n_s}}K_1(M_{\pi} L\sqrt{n_s})
\notag\\
&
+\frac{2\big((m_\Delta-m)(m_\Delta+m)^3-2(m_\Delta^2-3m m_\Delta-2m^2)M_{\pi}^2 + M_{\pi}^4\big)}{M_\pi L\sqrt{n_s}}
K_1(M_{\pi} L\sqrt{n_s})
\notag\\
&
+\big((m_\Delta-m)(m_\Delta+m)^3-2(m_\Delta^2-3m m_\Delta-2m^2)M_{\pi}^2 + M_{\pi}^4\big)
\big(K_0(M_{\pi} L\sqrt{n_s})+K_2(M_{\pi} L\sqrt{n_s})\big)
\notag\\
&
-\frac{16 m^2}{L^2 n_s}K_2(M_{\pi} L\sqrt{n_s})
+\frac{4m^2 M_{\pi}}{L\sqrt{n_s}}
\big(K_1(M_{\pi} L\sqrt{n_s})+K_3(M_{\pi} L\sqrt{n_s})\big)
\bigg\}
\ ,
\end{align}
with $\mathcal{M}^2_\Delta=L^2 n_s[x_1(m_\Delta^2+(x_1-1)m^2)-(x_1-1)M_{\pi}^2]$. Here $m_\Delta$ and $h_A$ are the mass of the $\Delta$ resonance and the LO $\pi N\Delta$ coupling constant (see e.g. Ref.~\cite{Yao:2016vbz} for definition), respectively.

The total FVC to $\sigma_{\pi N}$ is given by
\begin{align}
\Delta_L  \sigma_{\pi N} = \Delta_L \sigma_{\pi N}^{(N)} (M_{\pi};L) + \Delta_L \sigma_{\pi N}^{(\Delta)} (M_{\pi};L)\ .\label{eq.deltaL.chpt.deltafull}
\end{align}
Since the $\Delta$ is strongly coupled to the $\pi N$ system, the second term can be regarded as partially incorporating two-loop contributions.

\begin{table}[tb]
\renewcommand{\arraystretch}{1.5}
\caption{FVC to the lattice QCD data of $\sigma_{\pi N}$ (in units of MeV) from Ref.~\cite{Agadjanov:2023efe}.}
\label{tab:FVC_lattice}
\centering
\begin{tabular}{c|r r r r r r r r r r r r r r r r}
\hline\hline
Ensemble    & H102 & N101 &  H105 & C101  & S400 & N451 & D450 & D452 &  N203 & S201 & N200 & D200 & E250 & N302 & J303 & E300\\
\hline
BChPT w/o $\Delta$ & $-4.72$ & $-0.94$ & $-6.17$ & $-1.49$ & $-8.22$ & $-1.70$ & $-0.68$ & $-1.11$ & $-2.86$ & $-16.3$ & $-3.90$ & $-1.73$ & $-0.52$ & $-9.07$ & $-3.80$ & $-1.06$\\
BChPT w $\Delta$  &  $-9.17$ & $-1.62$ & $-11.7$ & $-2.52$ & $-16.5$ & $-3.03$ & $-1.11$ & $-1.75$ & $-5.37$ & $-33.3$ & $-7.25$ & $-2.89$ & $-0.76$ & $-18.3$ & $-6.94$ & $-1.69$\\     
\hline\hline
\end{tabular}
\end{table}

To proceed, we use Eq.~\eqref{eq.deltaL.chpt.deltafull} to evaluate the FVC for the lattice QCD data implemented in our fit. The FVC values for each ensemble from Ref.~\cite{Agadjanov:2023efe} are compiled in Table~\ref{tab:FVC_lattice}. In our numerical estimate, we set $h_A=1.42$~\cite{Yao:2016vbz}, $m_\Delta=1.232$~GeV~\cite{ParticleDataGroup:2024cfk}, $F=86.7$~MeV~\cite{FlavourLatticeAveragingGroupFLAG:2024oxs}, $g_A=1.13$~\cite{Yao:2017fym} and $m=0.938$~GeV~\cite{ParticleDataGroup:2024cfk}. For the ensembles in Table~\ref{tab:FVC_lattice}, the dimensionless product $M_{\pi}L$ takes values in the interval $[3, 5.83]$ with $M_{\pi}$ ranging from $128$ to $352$~MeV. When $M_{\pi}L$ is small while $M_{\pi}$ remains large, the FVC becomes significant. This explains why ensemble S201 (with $M_{\pi}L=3$ and $M_{\pi}=288$ MeV) exhibits the largest FVC among all configurations. As shown in Table~\ref{tab:FVC_lattice}, the $\pi\Delta$ loop provides a sizable contribution to the FVC. In this way, we determine the FVC values for the lattice QCD data points beforehand, rather than determining them through fitting as done in Ref.~\cite{Agadjanov:2023efe}.

\subsection{Comparative fits}
We perform fits by minimizing the following $\chi^2$ function:
\begin{align}
\chi^2 = \chi^2_{m_N}+\omega\cdot(\chi^2_{\rm win} + \chi^2_{\rm sum} + \chi^2_{\rm exp})\ ,\quad \omega= 1/3\ ,
\end{align}
where
\begin{align}
   \chi^2_{m_N}=\frac{(m_N^{\rm ChPT}-m_N^{\rm phy.})^2}{(\delta m_N^{\rm phy.})^2} \ ,\quad  
   \chi^2_{\rm method}=\frac{(\sigma_{\pi N}^{\rm ChPT}-\sigma_{\pi N}^{\rm LQCD})^2}{(\delta \sigma_{\pi N}^{\rm LQCD})^2}\ ,\quad \textrm{method}\in\{\textrm{win},\textrm{sum},\textrm{exp}\}\ ,
\end{align}
with $m_N^{\rm phy.}\pm \delta m_N^{\rm phy.}=(938.92\pm 0.65)$~MeV~\cite{ParticleDataGroup:2024cfk}. The central value of the physical nucleon mass is taken as the average mass of the proton and neutron, while its uncertainty is set as one-half of their mass difference. The lattice QCD data are taken from Ref.~\cite{Agadjanov:2023efe}, containing three datasets (labeled by ``win", ``sum" and ``exp", respectively) obtained by using different methods of treating excited-state contributions. BChPT results of the nucleon mass and pion-nucleon sigma term are denoted by $m_N^{\rm ChPT}$ and $\sigma_{\pi N}^{\rm ChPT}$, respectively. For comparison, three different fits are tried by imposing chiral expressions up to $\mathcal{O}(p^3)$ (Fit-I), $\mathcal{O}(p^4)$ (Fit-II) and $\mathcal{O}(p^5)$ (Fit-III), respectively. Results of the fits are collected in Table~\ref{tab:fit.LECs.com}.

\begin{table}[tb]
\renewcommand{\arraystretch}{1.5}
\caption{Fit results of the two-loop renormalized LECs and the parameter $b_\pi$ characterizing the lattice spacing effect. The last two rows show the fit quality $\chi^2/{\rm d.o.f.}$ and the prediction of $\sigma_{\pi N}$ ($\bar{\sigma}_{\pi N}$) at the physical charged (neutral) pion mass, respectively.
}
\label{tab:fit.LECs.com}
\begin{tabular*}{\textwidth}{@{\extracolsep{\fill}}lccc}
\hline\hline
{LECs} & Fit-I: $\mathcal{O}(p^3)$  & Fit-II: $\mathcal{O}(p^4)$  & Fit-III: $\mathcal{O}(p^5)$ 
\\
\hline
$\tilde{m}$~ [MeV]  & $886.7\pm 2.0$ & $872.9\pm 1.9$ & $863.9\pm 2.1$ \\
$\tilde{c}_1$~ [GeV$^{-1}$] & $-0.84\pm0.02$ &  $-1.09\pm0.02$ & $-1.06\pm0.02$ \\
$\tilde{e}_m$  [GeV$^{-3}$] & - & $-1.34\pm 0.21$    & $-5.59\pm 0.21$\\
$b_\pi$ [GeV$^{-1}$] &$0.18\pm 0.20$ & $0.20\pm 0.20$ & $0.20\pm 0.20$  \\
\hline
${\chi^2}/{\rm d.o.f.}$  & {$\frac{21.5}{(15+1-3)}\simeq 1.65$}& {$\frac{18.83}{(15+1-4)}\simeq 1.57$} & {$\frac{18.24}{(15+1-4)}\simeq 1.52$}\\
\hline
$\sigma_{\pi N}$~[MeV] & $48.1\pm 1.9$ &$56.7\pm 1.9 $     &$55.9\pm 1.9$ \\
$\bar{\sigma}_{\pi N}$~[MeV] &$45.4\pm 1.8$&$53.9\pm 1.8 $& $53.1\pm 1.8$ \\
\hline\hline
\end{tabular*}
\end{table}
\end{document}